\documentclass[journal]{IEEEtran}
\usepackage[latin1]{inputenc}
\usepackage{graphicx}
\usepackage{tikz}
\usepackage{amsmath}
\usepackage{amsfonts}
\usepackage{amssymb}
\usepackage{amsthm}
\usepackage[sans]{dsfont}
\usepackage{verbatim}
\usepackage{float}
\usepackage{stfloats}
\usepackage{topcapt}
\usepackage{xcolor}
\usepackage{threeparttable}
\restylefloat{table}

\newcommand{\hv}{{\bf h}}

\newcommand{\rv}{{\bf r}}
\newcommand{\sv}{{\bf s}}

\newcommand{\wv}{{\bf w}}

\newcommand{\zv}{{\bf z}}
\newcommand{\zerov}{{\bf 0}}

\newcommand{\Am}{{\bf A}}

\newcommand{\Hm}{{\bf H}}

\newcommand{\Om}{{\bf O}}

\newcommand{\Rm}{{\bf R}}

\newcommand{\Wm}{{\bf W}}

\newcommand{\Xm}{{\bf X}}

\newcommand{\Ac}{{\cal A}}

\newcommand{\Cc}{{\cal C}}

\newcommand{\Nc}{{\cal N}}
\newcommand{\Oc}{{\cal O}}

\newcommand{\Hcb}{\pmb{\cal H}}

\allowdisplaybreaks

\begin{document}
\title{A New Low-Complexity Decodable Rate-1 Full-Diversity $4 \times 4$ STBC with Nonvanishing Determinants}
\author{\IEEEauthorblockN{Amr~Ismail,~\IEEEmembership{IEEE Student Member} Jocelyn~Fiorina,~\IEEEmembership{IEEE Member} and~Hikmet~Sari,~\IEEEmembership{IEEE Fellow}}\\
\IEEEauthorblockA{Telecommunications Department, SUPELEC, F-91192 Gif-sur-Yvette, France\\
Email:$\lbrace$amr.ismail, jocelyn.fiorina, and hikmet.sari$\rbrace$@supelec.fr\\}}

\maketitle
\begin{abstract}
Space-time coding techniques have become common-place in wireless communication standards  as they provide an effective way to mitigate the fading phenomena inherent in wireless channels. However, the use of Space-Time Block Codes (STBCs) increases significantly the optimal detection complexity at the receiver unless the low complexity decodability property is taken into consideration in the STBC design. In this letter we propose a new low-complexity decodable rate-1 full-diversity $4\times 4$ STBC. We provide an analytical proof that the proposed code has the Non-Vanishing-Determinant (NVD) property, a property that can be exploited through the use of adaptive modulation which changes the transmission rate according to the wireless channel quality. We compare the proposed code to existing low-complexity decodable rate-1 full-diversity  $4\times 4$ STBCs in terms of performance over quasi-static Rayleigh fading channels, detection complexity and Peak-to-Average Power Ratio (PAPR). Our code is found to provide the best performance and the smallest PAPR which is that of the used QAM constellation at the expense of a slight increase in detection complexity w.r.t. certain previous codes but this will only penalize the proposed code for high-order QAM constellations.
\end{abstract}


\section{Introduction}
\IEEEPARstart{S}{pace}-time coding techniques have become common-place in wireless communication standards \cite{802.11n,802.16e} as they provide an effective way to mitigate the fading phenomena inherent in wireless channels. However, the use of Space-Time Block Codes (STBCs) increases significantly the optimal detection complexity at the receiver unless the low-complexity decodability property is taken into consideration in the STBC design. By complexity of detection, we mean the minimum number of times the receiver has to compute the Maximum Likelihood (ML) metric to estimate the transmitted symbols \cite{biglieri}. It is well known that Complex Orthogonal Design (COD) codes \cite{tarokh,tirkkonen} are Single-Symbol Decodable (SSD) for general constellations. But if one considers the case of rectangular QAM constellations, the detection of each complex symbol reduces to separate detection of two real symbols which can be effectively performed via two threshold detectors (or equivalently PAM 
s) with a fixed complexity that does not depend on the constellation size. On the other hand, the rate of square COD codes decreases exponentially with the number of transmit antennas \cite{tirkkonen}, which makes them more suitable for low-rate communications. In this letter we address the issue of increasing the rate of the COD at the expense of an increase in the detection complexity while preserving the coding gain and the Non-Vanishing-Determinant (NVD) property \cite{belfiore}. This property can be exploited through the use of adaptive modulation which varies the transmission rate (through the choice of the modulation order) according to the wireless channel quality. The decoder exploits the STBC structure by estimating the set of the $n_1$ orthogonal symbols assuming the knowledge of the other set of $n_2$ symbols, then performing the ML exhaustive search over $n_2$ symbols which reduces the complexity of detection from $M^{\left( n_1+n_2\right)}$ to $M^{n_2}$ where $M$ is the rectangular QAM constellation size. Based on this conditional ML strategy \cite{serdar}, we construct a new low-complexity decodable rate-1 full-diversity $4 \times 4$ code that encloses the rate-$3/4$ COD code in \cite{tirkkonen}. Furthermore, we provide an analytical proof (see Appendix) that the proposed code has the NVD property thus achieving the diversity-multiplexing trade-off \cite{elia}. 

We compare the proposed code to existing low-complexity decodable rate-1 full-diversity STBCs in terms of performance over quasi-static Rayleigh fading channels, detection complexity and Peak-to-Average Power Ratio (PAPR). Our code is found to provide the best performance, the smallest PAPR, which is that of the used QAM constellation, at the expense of a slight increase in detection complexity w.r.t. certain previous codes only for large constellations.

The letter is organized as follows: In Section II, preliminaries on STBCs are provided. In Section III we introduce the new code and derive its low-complexity decodability property. In Section IV, performance comparison by means of computer simulations of the proposed code and known similar codes in the literature is presented. We give our conclusions in section V, and the proof of the NVD property for the proposed code is provided in the Appendix. 

\begin{table*}[b!]
\setcounter{equation}{11}
\hrule
\vspace*{2.4mm}
\begin{equation}
x^{\text{ML}}_i=\textit{sign}\left(\Re\lbrace\tilde{\hv}^H_i\rv\rbrace \right)\times \text{min}\left[\Big\vert 2 \ \textit{round}\left[\left(\frac{\Re\lbrace\tilde{\hv}^H_i\rv\rbrace}{\Vert\textit{vec}(\Hm)\Vert^2}-1\right)/2\right]+1 \Big\vert,\sqrt{M}-1 \tnote{*}\right]\ \forall i=1,\ldots,2K  
\label{slicer}
\end{equation}
\end{table*}

\subsection*{Notations:}
Hereafter, small letters , bold small letters and bold capital letters will designate scalars, vectors and matrices respectively. If $\Am$ is a matrix, then $\Am^H$, $\Am^T$ and $\textit{det}[\Am]$ denote the hermitian, the transpose and the determinant of $\Am$ respectively. We define the $\textit{vec}(.)$ as the operator which, when applied to  a $m \times n$ matrix, transforms it  into a $mn\times 1$ vector by simply concatenating vertically the columns of the corresponding matrix. The $\otimes$ operator is the Kronecker product and $\delta_{kj}$ is the kronecker delta. The $\textit{sign}(.)$ operator returns 1 if its scalar input is $\geq 0$ and -1 otherwise. The $\textit{round}(.)$ operator rounds its argument to the nearest integer. The $\Re(.)$ and $\Im(.)$ operators denote the real and imaginary parts, respectively, of their argument. For any two integers $a\ \text{and}\ b,\ a\equiv b \left(\text{mod}\ n\right)$ means that $a-b$ is a multiple of $n$.   
\setcounter{equation}{0}
\section{Preliminaries}
We define the MIMO channel input-output relation as: 
\begin{equation}
\underset{T\times N_R}{\Rm} =\underset{T\times N_T}{\Xm} \underset{N_T\times N_R}{\Hm} +\underset{T\times N_R}{\Wm}
\label{model}  
\end{equation} 
where $T$ is the number of channel uses, $N_R$ is the number of receive antennas, $N_T$ is the number of transmit antennas, $\Rm$ is the received signal matrix, $\Xm$ is the code matrix, $\Hm$ is the channel matrix with entries $h_{kl} \sim \Cc \Nc(0,1)$ and $\Wm$ is the noise matrix with entries $w_{ij} \sim \Cc \Nc(0,N_{0} )$.
In the case of Linear Dispersion (LD) codes \cite{hassibi}, an STBC that encodes $2K$ real symbols is expressed as a linear combination of the transmitted symbols as:
\begin{equation}
\Xm=\sum^{2K}_{k=1} \pmb{\beta}_kx_k
\label{LD}
\end{equation}
with $x_k\in\mathbb{R}$, $K$ is the number of complex transmitted symbols and the $\pmb{\beta}_k, k=1,...,2K$ are $T \times N_T$ complex matrices called dispersion or weight matrices. 
For the LD codes, the MIMO channel model can be expressed in a useful manner as shown below. Expanding the code matrix $\Xm$ in the channel model as in \eqref{LD}, we obtain:
\begin{equation}
\Rm=\sum^{2K}_{k=1}\left(\pmb{\beta}_k\Hm\right)x_k+\Wm.
\end{equation} 
Applying the $\textit{vec}(.)$ operator to the R.H.S and the L.H.S we obtain:
\begin{equation}
\textit{vec}(\Rm)=\sum^{2K}_{k=1}\left(I_{N_R}\otimes\pmb{\beta}_k\right)\textit{vec}\left(\Hm\right)x_k+\textit{vec}(\Wm).
\label{vec}
\end{equation}
where $I_{N_R}$ is the $N_R\times N_R$ identity matrix.\\
If $\rv_i$, $\hv_i$ and $\wv_i$ designate the ith column of the received signal matrix $\Rm$, the channel matrix $\Hm$ and the noise matrix $\Wm$ respectively, then equation \eqref{vec} can be written in  matrix form as :
\begin{equation}
\underbrace{\begin{bmatrix}\rv_1\\\vdots\\\rv_{N_R}\end{bmatrix}}_{\rv}=\underbrace{\begin{bmatrix}\pmb{\beta}_{1}\hv_1&\dots&\pmb{\beta}_{2K}\hv_1\\\vdots&\vdots&\vdots\\\pmb{\beta}_{1}\hv_{N_R}&\dots&\pmb{\beta}_{2K}\hv_{N_R}\end{bmatrix}}_{\Hcb}\underbrace{\begin{bmatrix}x_1\\\vdots\\x_{2K}\end{bmatrix}}_{\sv}
+\underbrace{\begin{bmatrix}\wv_1\\\vdots\\\wv_{N_R}\end{bmatrix}}_{\wv}.
\end{equation}
Thus we have:
\begin{equation}
\rv=\Hcb\sv+\wv
\label{model2}
\end{equation}
For COD codes, the weight matrices satisfy \cite{tirkkonen}: 
\begin{equation}
\pmb{\beta}^H_i\pmb{\beta}_j+\pmb{\beta}_j^H\pmb{\beta}_i=2\delta_{ij}I_{N_T}
\label{clifford}
\end{equation} 
In this case, if $\tilde{\hv}_i$ is the $i$'th column of the equivalent channel matrix $\Hcb$ in \eqref{model2}, we may write:
\begin{equation*} 
\begin{split}
\tilde{\hv}_k^H\tilde{\hv}_l+\tilde{\hv}_l^H\tilde{\hv}_k&=\textit{vec}(\Hm)^H\Big[(I_{N_R}\otimes\pmb{\beta}_k)^H(I_{N_R}\otimes\pmb{\beta}_l)\\
&+(I_{N_R}\otimes\pmb{\beta}_l)^H(I_{N_R}\otimes\pmb{\beta}_k)\Big]\textit{vec}(\Hm)\\
&=\textit{vec}(\Hm)^H\left[I_{N_R}\otimes(\pmb{\beta}^H_k\pmb{\beta}_l
+\pmb{\beta}^H_l\pmb{\beta}_k)\right]\textit{vec}(\Hm)\\
&=0,\ \forall\ k\neq l
\end{split}
\end{equation*}
where the last equality comes from \eqref{clifford}. Thus we have:
\begin{equation}
\Re\left\{\tilde{\hv}_k^H\tilde{\hv}_l\right\}=0\ \forall\ k\neq l.
\label{orthogonal}
\end{equation}
This relation can be used for the detection of orthogonal symbols as below. Expanding the MIMO channel model in \eqref{model2} we have:
\begin{equation}
\rv=\Hcb\sv+\wv=\sum^{2K}_{k=1}\tilde{\hv}_kx_k+\wv.
\end{equation}  
Multiplying both sides by $\tilde{\hv}^H_l$ we obtain:
\begin{equation}
\tilde{\hv}^H_l\rv=\Vert\tilde{\hv}_l\Vert^2x_l
+\underset{k\neq l}{\sum^{2K}_{k=1}}\tilde{\hv}^H_l\tilde{\hv}_kx_k+\tilde{\hv}^H_l\wv.
\end{equation}
Taking the real parts of both sides, the second element of the R.H.S vanishes thanks to \eqref{orthogonal} and we obtain:
\begin{equation}
\Re\left\{\tilde{\hv}^H_l\rv\right\}=\Vert\textit{vec}(\Hm)\Vert^2x_l+\underbrace{\Re\left\{\tilde{\hv}^H_l\wv\right\}}_{\tilde{w}_l}
\label{siso}
\end{equation}
where $\tilde{w}_l$ is a zero-mean real Gaussian noise. If we assume that the complex symbols are drawn from a square constellation such as 4-/16-QAM constellations, which is usually the case in practical scenarios, the real symbols may be decoded independently through a PAM slicer as illustrated in \eqref{slicer}.
It is worth noting that the PAM slicer equations require only a fixed number of simple arithmetic operations, which does not grow with the size of the square QAM constellation, and therefore they are considered of complexity $\Oc(1)$.
\setcounter{equation}{12}
\section{The Proposed Code}
\begin{figure}[h!]
      \centerline{\includegraphics[scale=0.58]{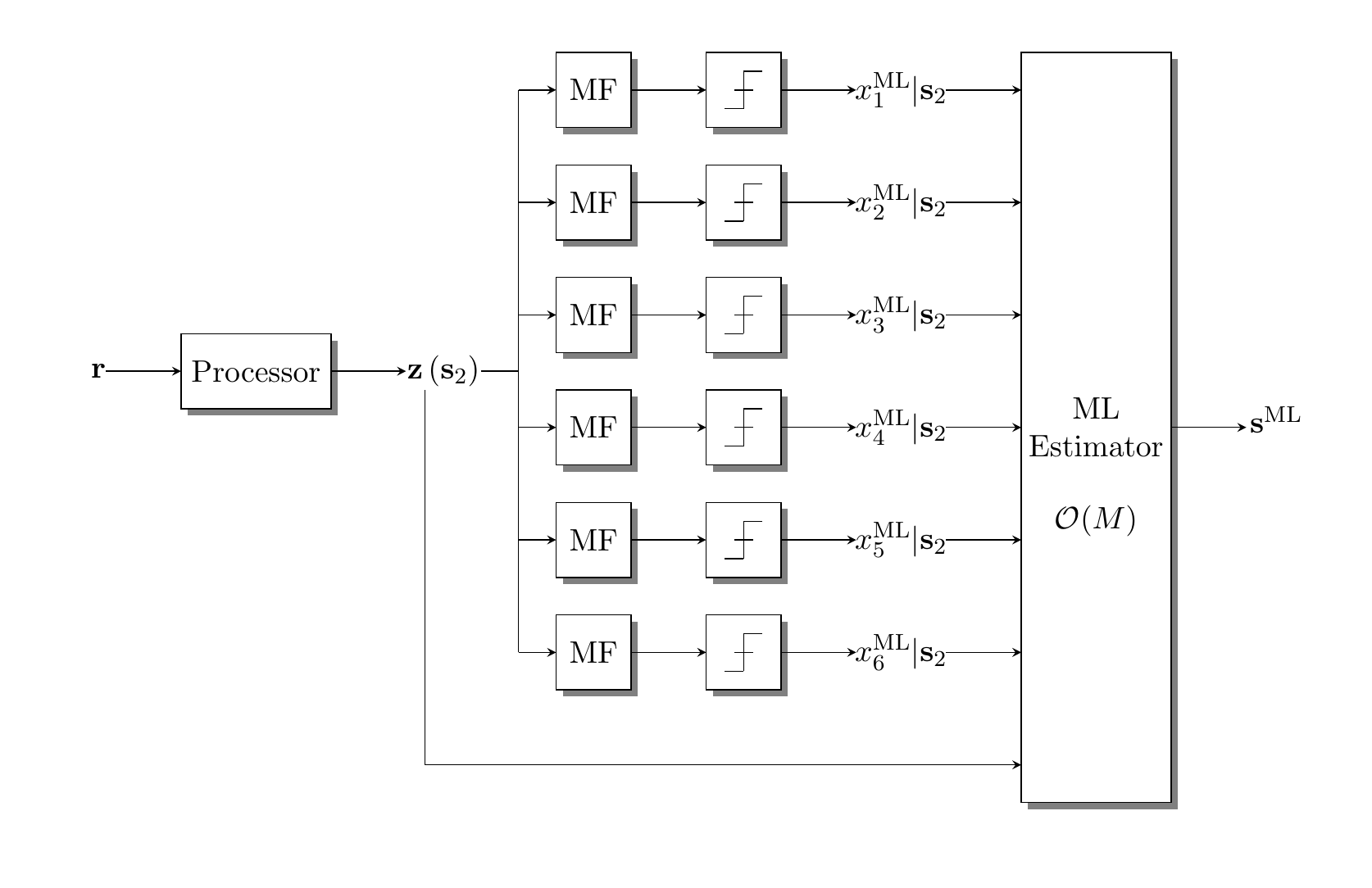} } 
      \caption{The low-complexity decoder of the proposed code} 
\label{dec} 
\end{figure}

From the previous section, a judicious structure for a new high-rate, low-complexity code would be to enclose a COD code into a new higher rate STBC. The resulting code will enjoy the low-complexity decodability through the use of conditional ML detection.\\
The proposed code denoted $\Xm_{\textit{new}}$ takes the following form:
\begin{equation}
\Xm_{\textit{new}}(\sv)=\Om(x_1,\ldots, x_6)+e^{j\phi}\left(\Rm_2 x_7+\Rm_3 x_8\right)\Rm_4 
\end{equation}
where the $\Om(x_1,\ldots,x_{2(a+1)})$ matrix denotes the square COD code for the case of $2^a$ transmit antennas and $\Rm_i$ is the matrix representations of the Clifford algebra generator $\gamma_i$ with $1\leq i \leq 2a+1$, $\sv=[\sv_1,\sv_2]$, $\sv_1=[x_1,\ldots,x_6]$, $\sv_2=[x_7,x_8]$ and $\phi$ is chosen to maximize the coding gain. The code matrix can be explicitly written as in \eqref{newcode}.
\begin{table*}[t!]
\begin{equation}
\Xm_{\textit{\textit{new}}}(\sv)=\begin{bmatrix} 
x_1+jx_2&x_3+jx_4&x_5+jx_6&-e^{j\phi}(x_7+jx_8)\\
-x_3+jx_4&x_1-jx_2&e^{j\phi}(-x_7+jx_8)&-x_5-jx_6\\
-x_5+jx_6&e^{j\phi}(x_7+jx_8)&x_1-jx_2&x_3+jx_4\\
-e^{j\phi}(-x_7+jx_8)&x_5-jx_6&-x_3+jx_4&x_1+jx_2
\end{bmatrix}
\label{newcode}
\end{equation}
\hrule
\end{table*}
It was found through exhaustive computer search that taking $\phi=\frac{1}{2}\cos^{-1}(1/5)$ maximizes the coding gain and that it remains constant up to 64-QAM unnormalized constellations. An analytical proof that the code in \eqref{newcode} possesses the NVD property is provided in Appendix. The decoder exploits the structure of the proposed code by computing the ML estimates of the orthogonal symbols $(x_1,\ldots, x_6)$ assuming that a given value of $(x_7,x_8)$ is transmitted which we will denote $\sv^{\text{ML}}_1\vert \hat{\sv}_2$. To do so, the decoder scans all of the possible values of $\sv_2$ and assigns to each of these values a new signal $\zv(\hat{\sv}_2)$:
\begin{equation}
\zv(\hat{\sv}_2)=\rv-\sum^{8}_{k=7}\tilde{\hv}_k\hat{x}_k
\end{equation}
In the case of the square QAM constellations, the ML estimates of the orthogonal symbols $(x_1,\ldots, x_6)$ assuming the knowledge of $(x_7,x_8)$ can be obtained exactly as in \eqref{slicer} with the only difference that $\rv$ will be replaced by $\zv(\hat{\sv}_2)$. The decoder computes then the ML metric for the current value of $(\sv^{\text{ML}}_1\vert \hat{\sv}_2,\hat{\sv}_2)$ according to:
\begin{equation}
D(\hat{\sv}_2)=\Big\Vert\zv(\hat{\sv}_2)-\sum^{6}_{i=1}\tilde{\hv}_ix^{\text{ML}}_i\Big\vert \hat{\sv}_2 \Big\Vert.
\end{equation}
The above metric calculation is repeated for every possible value of $\sv_2$ which means that the decoder makes $M$ metric computations. Finally, it decides in favor of the value of $\sv_2$ with the minimum ML metric:
\begin{equation}
\sv^{\text{ML}}_2=\text{arg}\ \underset{\hat{\sv}_2 \in \Ac}{\text{min}} D(\hat{\sv}_2)
\end{equation} 
where $\Ac$ is the set of the square QAM constellation points. The proposed decoder is illustrated in Fig.~\ref{dec}.

\section{Numerical and Simulations Results}
Table \ref{comparison} summarizes the comparison of the proposed code with existing similar low-complexity decodable codes in terms of detection complexity for square QAM constellations, Min \textit{det} and PAPR.
The minimum determinant is defined as:
\begin{equation}
 \text{Min}\ \textit{det}=\underset{\Delta\sv \in \Delta \Cc \backslash\lbrace \zerov \rbrace}{\text{min}}\vert\textit{det}\left[\left(\Xm(\Delta\sv) \right)\right]\vert=\sqrt{\delta} 
\end{equation}
where $\Delta \sv=\sv-\sv'$, $\Delta \Cc$ is the vector space spanned by $\Delta \sv$ and $\delta$ is the coding gain. For consistency, the minimun determinant is evaluated at fixed average transmitted power per channel use for all codes. The PAPR is defined as:
\begin{equation}
\text{PAPR}_n=\frac{\underset{t}{\text{max}}\vert\Xm(t,n)\vert^2}{T^{-1}\underset{t}{\sum} \mathbb{E}\lbrace\vert\Xm(t,n)\vert^2\rbrace}
\end{equation}
where $t \in \lbrace 1,\ldots T \rbrace$ and $n \in \lbrace 1,\ldots N_T \rbrace$. Due to the symmetry between transmit antennas, the subscript $n$ will be omitted. It is worth noting that the complexity of detection of the QOD code in \cite{papadias} can be reduced from $\Oc(M^2)$ to $\Oc(M)$ and the complexity of detection of the Single Symbol Decodable (SSD) codes in \cite{sinnokrot}-\cite{EAST} can be reduced from $\Oc(M)$ to $\Oc(\sqrt{M})$ as stated in \cite{thesis} by the use of conditional detection and PAM slicers as for the proposed code. 

\begin{figure}[h!]
      \centerline{\includegraphics[scale=0.7]{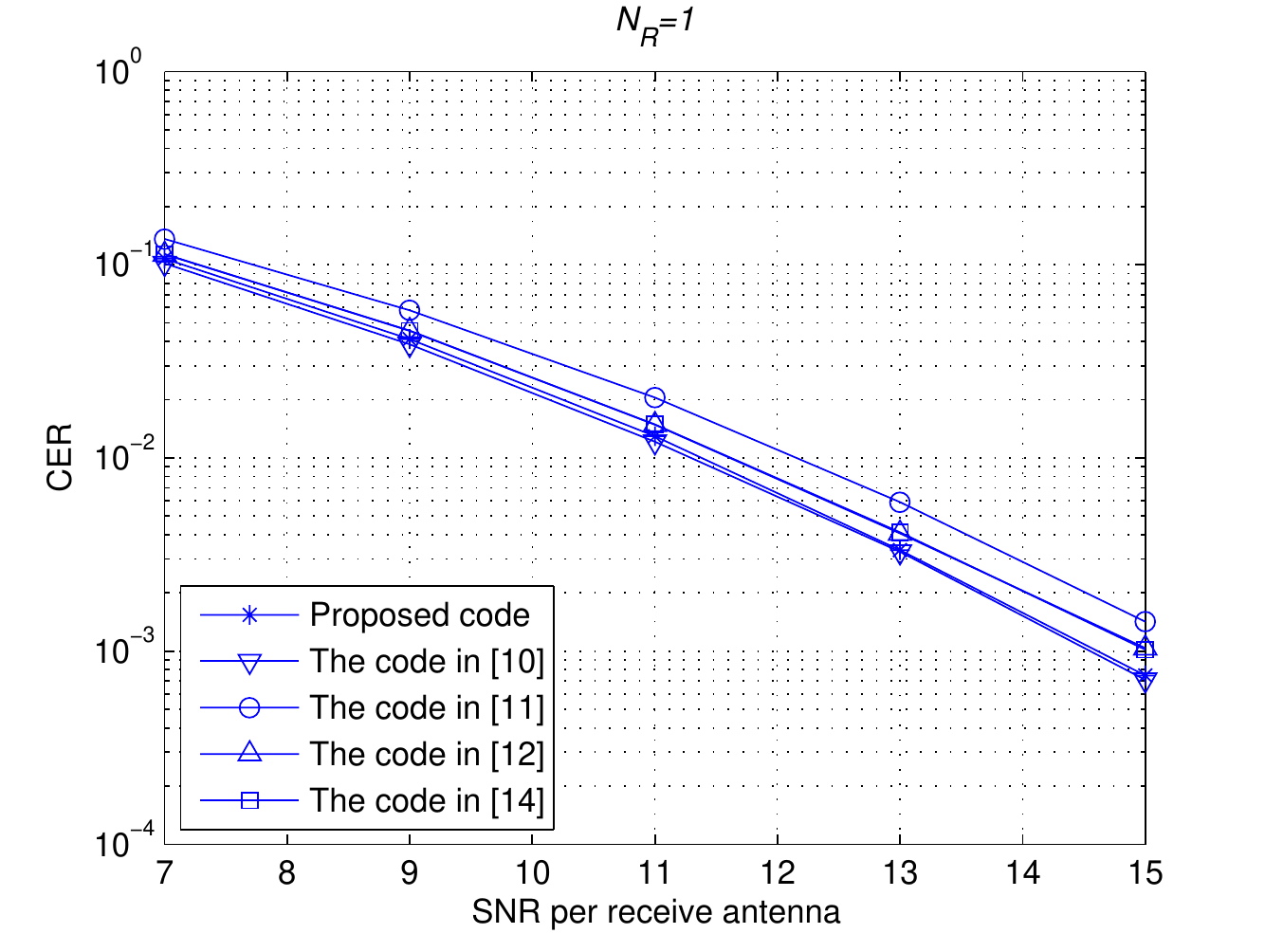}} 
      \caption{CER performance for $4\times 1$ configuration and 2 bpcu}
      \label{NR1}
\end{figure}

\begin{figure}[h!]
      \centerline{\includegraphics[scale=0.7]{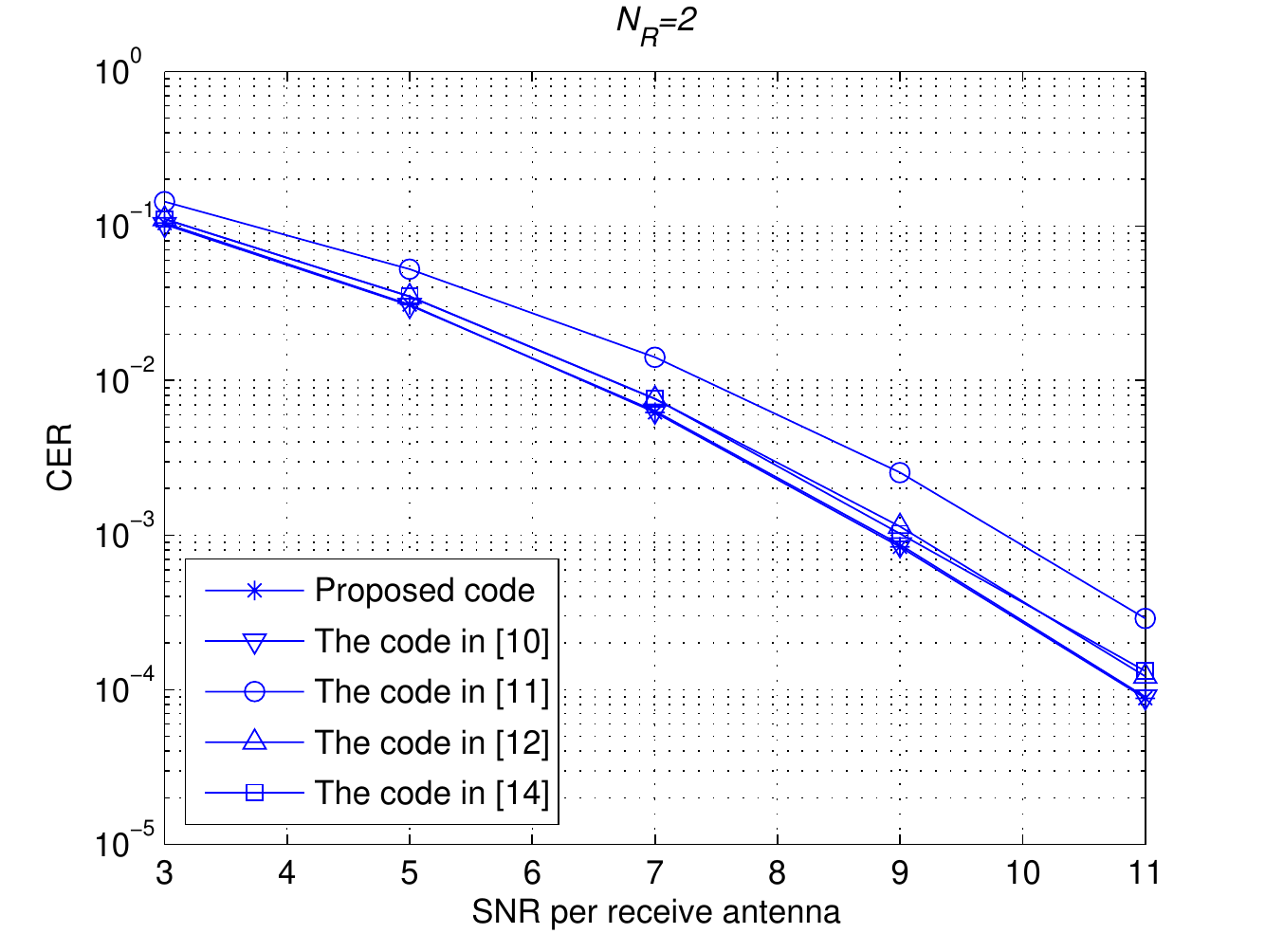}} 
      \caption{CER performance for $4\times 2$ configuration and 2 bpcu}
      \label{NR2}
\end{figure}

We can notice the following: The proposed code and the linear complexity decodable QOD code in \cite{papadias} achieve the highest coding gain compared to existing low-complexity decodable rate-1 full-diversity $4\times 4$ STBCs in the literature and thus we can expect that they provide better performance in the high SNR region which can be verified from Figs.~\ref{NR1},~\ref{NR2}. The proposed code along with the codes of \cite{papadias,sinnokrot} have the lowest PAPR which is that of the used constellation. On the other hand, our code suffers from a slight increase in decoding complexity w.r.t. the codes of \cite{sinnokrot}-\cite{EAST}, but this will only penalize the proposed code for high-order QAM constellations (in fact our code will be more complex to decode only for $M\geq 64$). The performance of the proposed code in terms of Codeword Error Rate (CER) versus SNR per receive antenna is compared with other low-complexity decodable rate-1 full-diversity STBCs in the literature. Simulations are carried out in a quasi-static Rayleigh fading channel in the presence of AWGN for $1$ and $2$ receive antennas and 2 bits per channel use (bpcu) (see Figs.~\ref{NR1},\ref{NR2}). One can easily verify that the proposed code and the Quasi-Orthogonal Design (QOD) code in \cite{papadias} provide the best performance. The rate-1 $4\times 4$ SAST, EOS and EAST codes in \cite{SAST,EOS,EAST} respectively, provide the same performance as the STBCs in \cite{SSD,CIOD} and therfore have been omitted to avoid congestion in Figs.~\ref{NR1},~\ref{NR2}.

\begin{table*}[t!]
\topcaption{summary of comparison in terms of complexity, Min \textit{det} and PAPR}
\centering
\begin{tabular}{|c|c|c|c|c|c|}
\hline
Code&Detection&Min \textit{det} ($=\sqrt{\delta}$)&\multicolumn{3}{c|}{PAPR (dB)}\\
\cline{4-6} &Complexity&for QAM constellations&QPSK&16QAM&64QAM\\
\hline
The proposed code&$M$&16&0&2.55&3.68\\
\hline
The QOD code in \cite{papadias}&$2M$&16&0&2.55&3.68\\
\hline
The SSD code in \cite{sinnokrot}&$4\sqrt{M}$&7.11&0&2.55&3.68\\
\hline
The codes in \cite{SSD,SAST}&$4\sqrt{M}$&12.8&1.61&4.16&5.29\\
\hline
The codes in \cite{CIOD,EOS,EAST}&$4\sqrt{M}$&12.8&5.79&8.34&9.47\\
\hline
\end{tabular}
\label{comparison}
\vspace*{2.4mm}
\hrule
\end{table*}

\section{Conclusion}
In the present letter, we have introduced a new low-complexity decodable rate-1 full-diversity $4\times4$ STBC that encloses the rate-$3/4$ COD and retains its coding gain. The coding gain is proved analytically to be constant over QAM constellations. Simulation results have shown that the performance of the proposed code is identical to the performance of the code in \cite{papadias} while being decodable at a lower complexity and it outperforms the codes in \cite{sinnokrot}-\cite{EAST} at the expense of a slight increase in decoding complexity for the $64$-QAM constellation and above. The proposed code has the same PAPR as the QOD code in \cite{papadias} and the code in \cite{sinnokrot}. 

\section*{Appendix}
In the following we will prove that choosing $\phi=\frac{1}{2}\cos^{-1}(1/5)$ indeed maximizes the coding gain $\delta$ (which is equal to $16^2$) for unnormalized QAM constellations and guarantees the NVD property for the proposed code. The coding gain $\delta$ is equal to the minimum Coding Gain Distance (CGD) \cite{jafarkhani}, or mathematically:
\begin{eqnarray}
\delta &=&\underset{\sv, \sv' \in \Cc}{\underset{\sv \neq \sv'}{\text{min}}} 
\underbrace{\textit{det}\left[\left(\Xm(\sv)-\Xm(\sv')\right)^H\left(\Xm(\sv)-\Xm(\sv')\right)\right]}_{\text{CGD}(\Xm(\sv),\Xm(\sv'))}\nonumber \\
&=&\underset{\Delta\sv \in \Delta \Cc \backslash\lbrace \zerov \rbrace}{\text{min}}
\vert\textit{det}\left[\left(\Xm(\Delta\sv) \right)\right]\vert^{2}
\end{eqnarray}
where $\Delta \sv=\sv-\sv'$, $\Delta \Cc$ is the vector space spanned by $\Delta \sv$. The code structure in \eqref{newcode} imposes:
\begin{equation*}
\underset{\Delta \sv \in \Delta \Cc \backslash\lbrace 0 \rbrace}{\text{min}}
\vert\textit{det}\left[\left(\Xm(\Delta \sv) \right)\right]\vert \leq \vert\textit{det}\left[ \Om(2,0,0,0,0,0\right]\vert=16
\end{equation*} 
As a result, the angle $\phi$ that maximizes the coding gain has to satisfy:
\begin{equation}
\vert\textit{det}\left[\left(\Xm(\Delta\sv) \right)\right]\vert \geq 16, \forall\ \Delta \sv \neq \zerov
\label{optimum}
\end{equation}
For the proposed code we have:
\begin{equation}
\begin{split}
\vert\textit{det}\left[\left(\Xm(\Delta\sv) \right)\right]\vert=&\Big\vert\left(\sum^{6}_{i=1}\Delta x_i^2 \right)^2+e^{j2\phi}b+e^{j4\phi}\left(\sum^{8}_{i=7}\Delta x_i^2\right)^2\Big\vert
\end{split}
\label{det}
\end{equation}
where:
\begin{equation*}
\begin{split}
b&=2\left(\sum^{6}_{i=1}\Delta x_i^2 \right)\left(\sum^{8}_{j=7}\Delta x_j^2\right)-4\Big[(\Delta x_7\Delta x_2-\Delta x_8\Delta x_6)^2\\
&+(\Delta x_7\Delta x_6+\Delta x_8\Delta x_2)^2+(\Delta x_7\Delta x_4-\Delta x_8\Delta x_3)^2\Big].
\end{split}
\end{equation*}
A simplification of the expression of $b$ is possible by noting that:
\begin{equation*}
\begin{split}
&\left(\sum^{6}_{i=1}\Delta x_i^2 \right)\left(\sum^{8}_{j=7}\Delta x_j^2\right)=
\left(\Delta x_2^2+\Delta x_6^2\right)\left(\Delta x_7^2+\Delta x_8^2\right)\\
&+\left(\Delta x_3^2+\Delta x_4^2\right) \left(\Delta x_7^2+\Delta x_8^2\right)
+\left(\Delta x_1^2+\Delta x_5^2\right) \left(\Delta x_7^2+\Delta x_8^2\right) .
\end{split}
\end{equation*}
Applying the Fibonacci's two square identities we obtain:
\begin{equation*}
\begin{split}
\left(\sum^{6}_{i=1}\Delta x_i^2 \right)\left(\sum^{8}_{j=7}\Delta x_j^2\right)
&=(\Delta x_7\Delta x_2-\Delta x_8\Delta x_6)^2\\
+(\Delta x_7\Delta x_6+\Delta x_8\Delta x_2)^2&+(\Delta x_7\Delta x_4-\Delta x_8\Delta x_3)^2\\
+(\Delta x_7 \Delta x_3+\Delta x_8\Delta x_4)^2&+(\Delta x_8\Delta x_1-\Delta x_7\Delta x_5)^2\\
+(\Delta x_7\Delta x_1+\Delta x_8\Delta x_5)^2&=\sum^{6}_{i=1}a_i^2,\ a_i=4\tilde{a}_i,\ \tilde{a}_i\in \mathbb{Z}.
\end{split}   
\end{equation*}
Therefore, one may write $b$ in more compact form a:
\begin{equation}
b=2\left(\sum^{6}_{i=1}a_i^2-2(a_1^2+a_2^2+a_3^2)\right) 
\label{b}
\end{equation}
Setting  $x=e^{j2\phi}$, the discriminant $\Delta$ of the second degree equation \eqref{det} is expressed as:
\begin{equation}
\Delta=4\left(\sum^{6}_{i=1}a_i^2-2\left(a_1^2+a_2^2+a_3^2\right)\right)^2-4\left(\sum^{6}_{i=1} a_i^2\right)^2\leq 0  
\end{equation}
Consequently, the roots of equation \eqref{det} are:
\begin{equation}
\lambda_{1,2}=\frac{-b\pm j\sqrt{4\left(\sum^{6}_{i=1}\Delta x_i^2\right)^2\left(\sum^{8}_{j=7}\Delta x_j^2\right)^2 -b^2}}{2 \left(\sum^{8}_{j=7}\Delta x_j^2\right)^2}
\end{equation}
where $\lambda_2=\lambda_1^*,\ \vert \lambda_1\vert= \vert \lambda_2 \vert =\frac{\sum^{6}_{i=1}\Delta x_i^2}{\sum^{8}_{j=7}\Delta x_j^2}$.

For the sake of simplicity, we will denote hereafter $\sum^{6}_{i=1}\Delta x_i^2=\sigma_1$ and $\sum^{8}_{j=7} \Delta x_j^2=\sigma_2$. In the case of $\sigma_1 \neq \sigma_2$, equation.~\ref{det} can be lower bounded as below:
\begin{equation}
\begin{split}
\vert\textit{det}\left[\left(\Xm(\Delta\sv)\right)\right]\vert\Big\vert_{\underset{\sigma_1\neq\sigma_2}{\Delta \sv \neq \zerov}}&=\sigma^2_2 \Big\vert \left(x-\lambda_1\right)\left(x-\lambda_2\right) \Big\vert\\ 
&\geq \sigma_2^2\Big\vert\left(\vert x\vert-\vert\lambda_1\vert\right)\left(\vert x\vert-\vert\lambda_2\vert\right)\Big\vert\\
&=\sigma_2^2\left(1-\frac{\sigma_1}{\sigma_2}\right)^2=\left(\sigma_2-\sigma_1\right)^2\geq 16
\end{split}
\label{ineqcase1}
\end{equation}
where the latter inequality follows by substituting $\Delta x_i=2n_i,\ n_i \in \mathbb{Z}$ as we are dealing with unnormalized QAM constellations. If $\sigma_1=\sigma_2=\sigma$, equation \eqref{det} can be written as
\begin{equation}
\vert\textit{det}\left[\left(\Xm(\Delta\sv) \right)\right]\vert\Big\vert_{\underset{\sigma_1=\sigma_2}{\Delta \sv \neq \zerov}}=\left\vert 2\sigma^4\cos(2\phi)+b\right\vert
\label{case2}
\end{equation}
where $b=2\left(\sigma^2-2(a_1^2+a_2^2+a_3^2)\right)$. Taking $\cos(2\phi)=1/5$, we have to prove that:
\begin{equation}
\Big\vert \frac{2\sigma^2}{5}+b\Big\vert \geq 16 \ \forall\ \sigma_1=\sigma_2=\sigma\neq 0.
\label{ineq1}
\end{equation}
Multiplying both sides by 5 and using \eqref{b}, the above inequality becomes:
\begin{equation}
\left\vert 12 \sigma^2-20(a_1^2+a_2^2+a_3^2)\right\vert_{\left(a_1,\ldots,a_6\right)\neq \zerov} \geq  5\times 16
\end{equation}
Denoting $\tilde{a}_i=\frac{a_i}{4}$ and $\tilde{\sigma}=\frac{\sigma}{4}$ we may have:
\begin{equation}
\left\vert 12 \tilde{\sigma}^2-20(\tilde{a}_1^2+\tilde{a}_2^2+\tilde{a}_3^2)\right\vert_{\left(\tilde{a}_1,\ldots,\tilde{a}_6\right)\neq \zerov} \geq  5.
\end{equation}
The above inequality is satisfied iff:
\begin{equation}
\left\vert 3 \tilde{\sigma}^2-5(\tilde{a}_1^2+\tilde{a}_2^2+\tilde{a}_3^2)\right\vert_{\left(\tilde{a}_1,\ldots,\tilde{a}_6\right)\neq \zerov} \geq  2.
\end{equation}
However, the L.H.S of the above inequality can be considered as a special case of:
\begin{equation}
3X^2_1-5(X^2_2+X^2_3+X^2_4)\Big\vert_{\left(X_1,X_2,X_3,X_4\right)\neq \zerov } 
\label{diophantine}
\end{equation}
where $X_i \in \mathbb{Z}$. This type of equations have been extensively studied in the mathematical literature dealing with  the solvability of quadratic Diophantine equations (see \cite{mordell}). Applying theorem 6  in \cite{mordell} we have:
\begin{equation}
3X^2_1-5(X^2_2+X^2_3+X^2_4) \neq 0\ \forall\ \left(X_1,X_2,X_3,X_4\right)\neq \zerov 
\end{equation} 
as $-3\times 5 \times 5\times 5\equiv 1\ (\text{mod}\ 8)$ with $3-5-5-5=-12\equiv 4\ (\text{mod}\ 8)$.\\
Moreover, 
\begin{equation}
3X^2_1-5(X^2_2+X^2_3+X^2_4)\neq\pm 1.
\end{equation}
Otherwise, we must have:
\begin{equation}
 3X^2_1\equiv \pm 1\ (\text{mod}\ 5)
\end{equation}
which cannot be true, since the quadratic residues modulo 5 are 0,1 and 4 \cite{quadres}, thus $3X^2_1 \equiv 0,\pm 3\  \text{or}\pm 2\ (\text{mod}\ 5)$. Therefore, we can write:
\begin{equation}
3X^2_1-5(X^2_2+X^2_3+X^2_4)\Big\vert_{\left(X_1,X_2,X_3,X_4\right)\neq \zerov} \geq 2 
\end{equation}
which in turns implies:
\begin{equation}
\vert\textit{det}\left[\left(\Xm(\Delta\sv) \right)\right]\vert\Big\vert_{\underset{\sigma_1=\sigma_2}{\Delta \sv \neq \zerov}}>16
\label{ineqcase2}
\end{equation}
thus ending the proof.

\end{document}